# Fluctuation-Enhanced Sensing of Bacterium Odors

Hung-Chih Chang[(+),] Laszlo B. Kish[(+)], Maria D. King[(x)], and Chiman Kwan[(*)]

[(+)]*Department of Electrical and Computer Engineering, Texas A&M University, College Station, TX 77843-3128, USA*

[(x)]*Department of Mechanical Engineering, Texas A&M University, College Station, TX 77843- 3123, USA*

[(*)]*Signal Processing, Inc., 13619 Valley Oak Circle, Rockville, MD 20850, USA*

**Abstract**

The goal of this paper is to explore the possibility to detect and identify bacteria by sensing their odor via fluctuation-enhanced sensing with commercial Taguchi sensors. The fluctuations of the electrical resistance during exposure to different bacterial odors, Escherichia coli and anthrax-surrogate Bacillus subtilis, have been measured and analyzed. In the present study, the simplest method, the measurement and analysis of power density spectra was used. The sensors were run in the normal heated and the sampling-and-hold working modes, respectively. The results indicate that Taguchi sensors used in these fluctuation-enhanced modes are effective tools of bacterium detection and identification even when they are utilizing only the power density spectrum of the stochastic sensor signal.

## 1. Introduction

Bio-defense, including bacterium detection, is one of the most important topics of defense to keep humans, animals and plants from the exposure of biologic warfare. Bacteria and their spores are popular biologic warfare agents for their low cost of fabrication, and the high damage potential on societies and economies. The rapid, and preferably on-site, evaluation and identification of airborne microorganisms is critical, especially in case of a bioterroristic threat. The currently available bacterial detection methods often require long culturing periods, expensive and bulky equipment and trained personnel. The ideal characteristics of a bacterial detection method are: it should be simple, practical, rapid, sensitive, specific, portable and inexpensive. One such a potential way of bacterium sensing is to analyze their odor [1][2]. Due to the low cost, wide availability, good sensitivity and selectivity, solid-state sensors, such as Taguchi sensors, could be such candidates of biologic sensors.

Taguchi sensors, which are commercially available, are heated semiconductor oxide films (usually $SnO_2$) with broad applications, including safety monitors for detecting combustible, pollution and toxic gases. The operation principle of the Taguchi sensors is based on the change of the sensor resistance because the gaseous agent diffuses into the film, breaks into molecular fragments, and at the grain boundaries it changes the conductivity of intergrain junctions by acting as donor or acceptor of electrons. Multiple gas identification can be done with single sensors [3-8] with temperature modulation but with limited selectivity [9,10]. To get better selectivity to identify gas mixtures arrays of different sensors are used. The typical devices are sensor systems combined with a pattern recognition unit, so-called "Electronic Nose [11-14]" (EN, for odors) or "Electronic Tongue" (ET, for liquid phase). Recently, ENs and ETs have been applied in various fields including environmental, agricultural and medical applications, and also in the food, beverage and automotive industries. The test personnel can use these devices instead of their own nose and tongue, thus avoid exposure to the agents and have instead a quantitative tool for detection and identification.

A specific way to enhance the sensitivity and selectivity of Taguchi and other gas sensors is to amplify, measure and analyze the small stochastic component of the sensor signal because the chemical environment usually influences these stochastic fluctuations. This way of operation is called Fluctuation-Enhanced Sensing (FES) [3-8,11,15-28]. The gas molecule fragments are executing a random walk along the grain boundary thus their conductance modulation effect is time dependent with a stochastic nature. The induced conductance noise has the fingerprint of the chemical agent because the doping properties and diffusion constant of the fragments are specific for each chemical agent.

For Taguchi sensors, two ways of FES is described:

*i.* Regular sensing (RS) method with heated sensor. In this case, the sensors are heated stationarily during the measurement. It is advised to stop the air flow for the duration of the data collection to avoid excess noises due to temperature fluctuations caused by the turbulence of air flow.

*ii.* Sampling-and-hold (SH) method. In this case, the sensor is heated for s short time, for a minute or so, (while the air can flow or not). Then the heating (and gas flow) is turned off and, after the sensor cooled down, the stochastic signal recording takes place. In this case, the gas fragments are trapped in the film and escape slowly, depending on the type of agent, so the measurement can also be done later. Another advantage of this method is that the noise induced by temperature fluctuations due to microscopic turbulence in the hot air convection is avoided.

In this work, the resistance fluctuations of three commercial gas sensors were used to identify three biological samples. There were two types of bacterium samples: Escherichia coli (Ecoli) and anthrax-surrogate Bacillus subtilis; and third type of sample was their culture medium, tryptic soy agar (TSA). We used the most conventional way of FES by measuring the power density spectrum (noise spectrum) of the stochastic signal component.

**2. Sample preparation**

The microorganisms used in this study were two harmless laboratory strains, *Escherichia coli* as a surrogate for pathogenic vegetative bacteria and *Bacillus globigii*, as a surrogate for pathogenic spores e.g. anthrax. Mid-log phase ($OD_{600}$ = 0.5, Optical density at 600 nm) cultures of *E. coli* K12 MG1655 (*E. coli* Genetic Resources at Yale CGSC, The Coli Genetic Stock Center, New Haven, NE) were grown in Luria Bertani (LB) medium [29] for about 4 hours at $37°C$ and at 150 RPM. One hundred microliters of the *E. coli* culture were spread on Difco Tryptic Soy Agar (TSA) plates (Becton Dickinson Co., Sparks, MD), and the plates were incubated overnight at $37°C$ [29].

To grow the spores of Anthrax surrogate, 50 mg of lyophilized Bacillus subtilis powder (U.S. Army Edgewood Proving Ground, Edgewood, MD) was re-suspended by vigorous vortexing in 5 mL of sterile deionized water and centrifuged at 4000

RPM for 9 min to remove traces of the culture medium. The supernatant was aspired and the pellet was resuspended in 10mL of sterile deionized water. One hundred microliters of the stock Bacillus subtilis were spread on TSA plates and incubated overnight at 30°C [14].

As reference, sterile TSA plates without bacteria were also prepared. As the TSA medium itself has a strong smell, identical amounts (27mL) of TSA medium were poured into each plastic Petri plate (VWR, Bridgeport, NJ) to maintain a constant level of background odor[14].

3. Experimental Setup

Three different types of commercial sensors SP11 (Figaro Inc.), SP32 (Figaro Inc.) and TGS2611 (FIS Inc.) are used in these experiments. These gas sensors are based on a specifically doped $SnO_2$ layer deposited on a ceramic substrate with a heater. The sensors SP32 and SP11 have a high sensitivity to hydrocarbons, hydrogen, alcohol and refrigerant gases. The sensor TGS2611 has a high sensitivity to organic solvent vapors including liquid petroleum gases, volatile organic compounds, etc. The new sensors were initially preheated ("burned-in") with the nominal heating voltage until they developed stable power spectrum, and this process took typically 5 days.

The measurement system is shown in Fig. 1. The three sensors are placed in the grounded stainless steel sensor chamber (volume 700 $cm^3$) where the odors generated by the samples can accumulate. These sensors are heated by a stable power supply (XP650) and the sensing resistor is driven by a low-noise DC current. The induced voltage fluctuations are amplified by an SR560 Low Noise Preamplifier. In the frequency domain, the power spectra of the amplified voltage noise across sensors were measured by an SR785 Dynamic Signal Analyzer. The power spectrum $S_u(f)$ of the voltage fluctuation was measured in the frequency range $100Hz\sim 100kHz$.

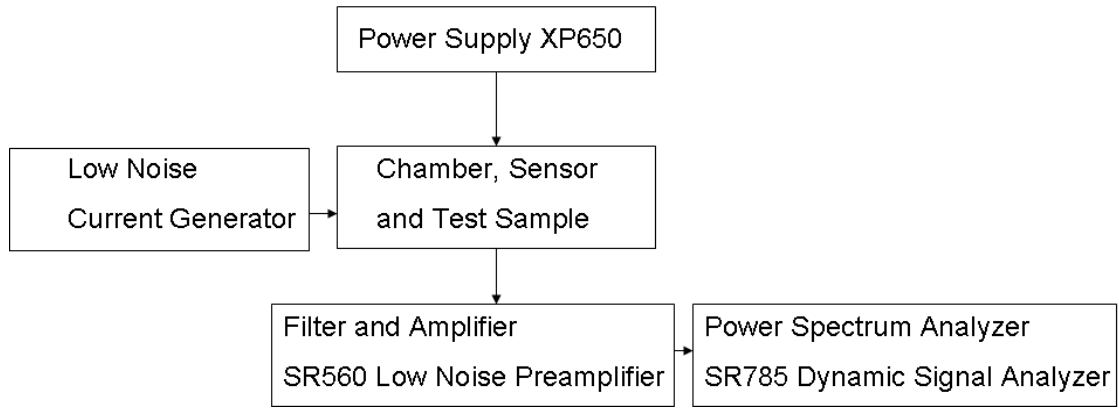

Figure 1. Fluctuation-enhanced odor sensing system.

## 4. Experimental procedure

For all test samples, the power spectra of all sensors over the measured range of frequencies scales roughly as $1/f$. The sensor resistance has been Ohmic in the observed range of DC voltage (0.3-6 V). The measured power density spectrum of voltage is proportional to the square of the DC voltage which confirm the resistance fluctuations origin of the voltage fluctuations [30][31].

To enhance the selectivity and sensitivity of sensors, all sensors were measured both by the stationarily heated (*heated*) FES [3,4] and the *sampling-and-hold* [5,6] FES methods. In the heated case, the nominal heating voltage (5V) is applied during the whole measurement. After the sample is placed in the chamber and the chamber is closed, we flush the chamber with synthetic air (for 3 minutes) and then we wait (typically 5 minutes) until a stable odor and the corresponding stable spectrum develops.

Each sampling-and-hold measurement [5,6] is preceded by a heated FES measurement sequence with stationary heating. Then, to execute the sampling-and-hold FES measurement, the heating is turned off while we are monitoring the spectrum. After 5 minutes, if the spectrum is stable, we record the spectrum which is the output pattern of the sampling-and-hold measurement [5,6].

Because the sensitivity of the system against resistance fluctuation depends on the value of the series resistance providing the battery-driven DC current drive for the sensor bias, whenever the measured voltage fluctuation were too small and close to the baseline, we changed these resistors for a proper one yielding sufficient large FES signals.

Four conditions have been tested in these experiments, empty chamber, TSA only, TSA + E. coli, and TSA + Anthrax-surrogate. After removing the sample, the chamber is flushed by synthetic air for 3 minutes. To see the reproducibility of the spectra, after completing the whole sequence of experiments with all the different samples, we repeated the whole sequence of tests with all the samples. In a few cases, we re-tested the samples two or four times.

From the measured he DC voltage on the sensor, the voltage of the battery, the value of the serial driver resistance, and the actual value of preamplification, we have calculated the calibrated the output data. These are the measured (raw) power density spectrum $S_u(f)$ of the FES voltage signal, which is circuit dependent; and the normalized power density spectrum $S_r(f)/R_s^2$, of the sensor resistance fluctuations which is generating the measured FES signal, where $R_s$ is the actual sensor resistance. An important advantage of evaluating and using $S_r(f)/R_s^2$ that it is independent from the actual circuit and bias and it is the characteristic of the sensor and the odor only. Thus $S_r(f)/R_s^2$ is much more suitable to compare measurement data obtained in different labs with different circuitry of settings.

The measurement circuitry is shown in Figures 2. The normalized power density spectrum $S_r(f)/R_s^2$, can be derived from the following equations.

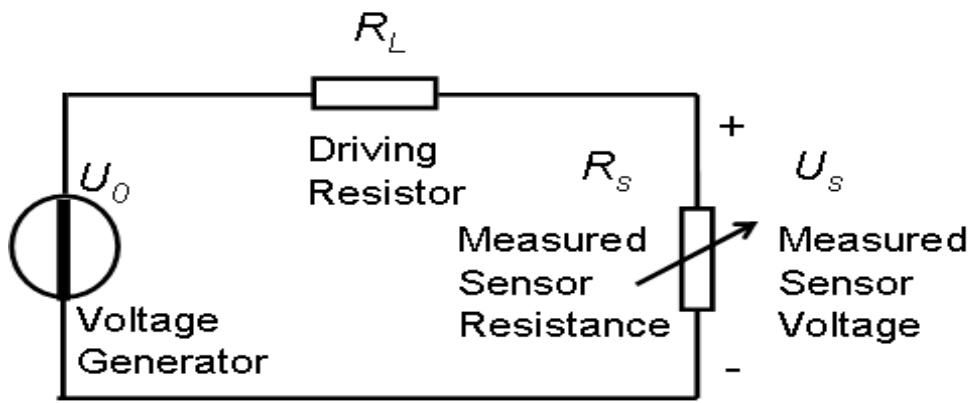

Figure 2. The measurement circuitry.

where $U_0$ is the driving DC voltage of voltage generator, $U_s$ is the measured DC voltage of sensor, $R_L = 36\,\text{k}\Omega$ is the serial resistor and $R_s$ is the actual resistance of the measured sensor.

The relationship between applied DC voltage $U_0$ and measured sensor voltage $U_s$ can be obtained from the relationship of the voltage divider:

$$U_s = U_0 \frac{R_s}{R_L + R_s} \quad . \tag{1}$$

Then the evaluation formula for the actual resistance of the measured sensor is

$$R_s = R_L \frac{U_s}{U_0 - U_s} \quad . \tag{2}$$

Furthermore, we define a parameter $\Delta$:

$$\Delta = \frac{dU_s}{dR_s} = U_0 \frac{R_L}{\left(R_L + R_s\right)^2} \quad . \tag{3}$$

Thus the relationship between $S_r(f)$, the mean-square resistance fluctuations $dR_s^2$ of the sensor in an infinitesimally small $df$ bandwidth, the measured mean-square voltage fluctuations in $df$ bandwidth, and the measured $S_u(f)$ are as follow:

$$S_r(f)df = dR_s^2 = \frac{dU_s^2}{\Delta^2} = \frac{S_u(f)}{\Delta^2} df \quad . \tag{4}$$

Thus:

$$S_r(f) = \frac{S_u(f)}{\Delta^2} = \frac{S_u(f)}{U_0^2} \left[\frac{\left(R_L + R_s\right)^2}{R_L}\right]^2 \tag{5}$$

In conclusion, the normalized power density spectrum can be determined from the measured voltage spectrum $S_u(f)$ with the following equation.

$$\frac{S_r(f)}{R_s^2} = \frac{S_u(f)}{U_0^2} \left[ \frac{(R_L + R_s)^2}{R_L R_s} \right]^2 \tag{6}$$

Finally, for the normalized plots in this paper, we multiply the normalized spectra with the frequency, following the usual practice [3-8, 18-23], in order to have a better naked-eye-visibility of the differences.

## 4. Experiment Results and Discussion

The raw power spectra of SP32, TGS 2611 and SP11, and the normalized spectra, for both the heated and sampling-and-hold sensor measurements, are shown in Figs.3-14. A good reproducibility and negligible memory effects are indicated by the fact that the spectra obtained with the same sensor in the empty chamber, before and after measurements with samples in the chamber, practically overlap each other. Similarly, the spectra of the same sample measured by the same sensor also show good reproducibility.

In Fig. 3, the raw spectra measured with the sensor SP32 in the heated mode are shown. The power spectra can be clearly divided into four groups just like the types of the samples: Empty, TSA, TSA + Anthrax-surrogate and TSA + E.coli. Therefore, the voltage fluctuation spectra of SP32 can differentiate all the four types of samples. Similar to the power spectra in Fig. 3, the normalized power spectra of the sensor resistance in Fig.4 can also be distinguished. Interestingly, the spectra of the sensor SP32 in the heated mode without normalization can be distinguished more clearly than the spectra with normalization.

In the sampling-and-hold working mode, most of the spectra (raw and normalized) obtained with the sensor SP32 are also well distinguishable, see Fig. 5, except the measurements with the two bacteria (E.coli and Anthrax-surrogate) yielding similar patterns. However, in this case by using the normalized spectra provides some distinguishability between the two bacteria, see Fig. 6.

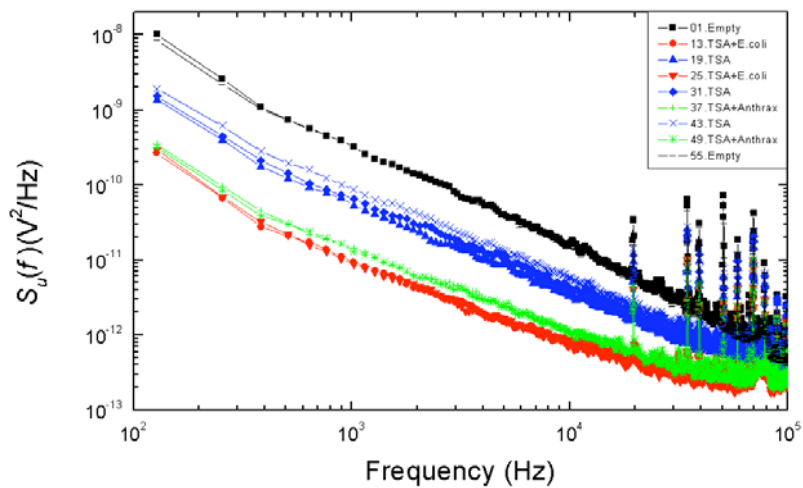

**Figure 3.** Raw Power Spectra of the Heated Sensor SP32. The alias "Anthrax" stands for anthrax surrogate *Bacillus subtilis*.

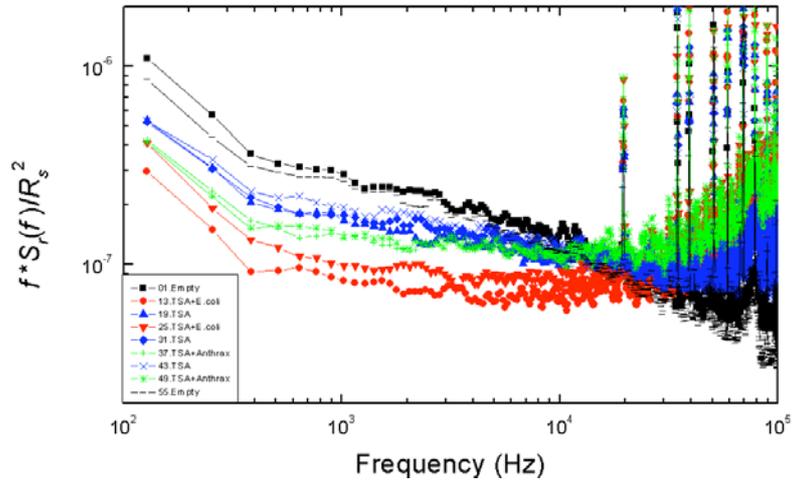

**Figure 4.** Normalized Power Spectra of the Heated Sensor SP32. The alias "Anthrax" stands for anthrax surrogate Bacillus subtilis.

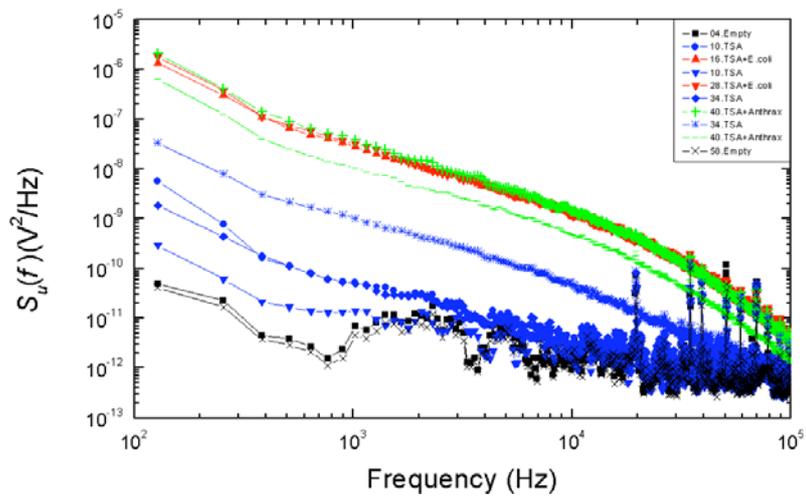

**Figure 5.** Raw Power Spectra of the Sampling-and-hold Sensor SP32. The alias "Anthrax" stands for anthrax surrogate *Bacillus subtilis*.

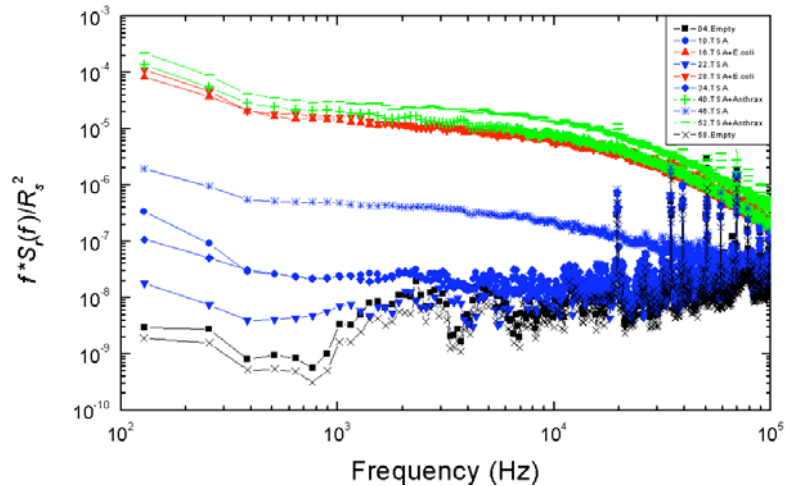

**Figure 6.** Normalized Power Spectra of the Sampling-and-hold Sensor SP32. The alias "Anthrax" stands for anthrax surrogate *Bacillus subtilis*.

In Fig. 7, the raw spectra measured with the sensor TGS 2611 in the heated mode are shown. The spectra cannot be clearly differentiated due to large variations and overlaps. However, the normalized spectra in the high-frequency limit ($f > 10\,\text{kHz}$) provides a sufficient separation between the four cases, see Fig. 8. Therefore, the normalized spectra of this sensor have better selectivity than the raw power spectra.

In the sampling-and-hold mode without and with normalization, see Figs. 9-10, the plot of the raw spectra of the sensor TGS2611 can clearly be divided into three groups: Empty and TSA; TSA + Anthrax-surrogate; and TSA + E. Coli, respectively. That means, the empty chamber and the chamber with the TSA are indistinguishable, which is fine for bacterium detection/identification. Compared with the result of heated sensor, the sampling-and-hold method can enhance the selectivity and sensitivity of this sensor. Besides, the spectra with normalization offer further enhancement of the differentiation, cf. Figs. 9 and 10.

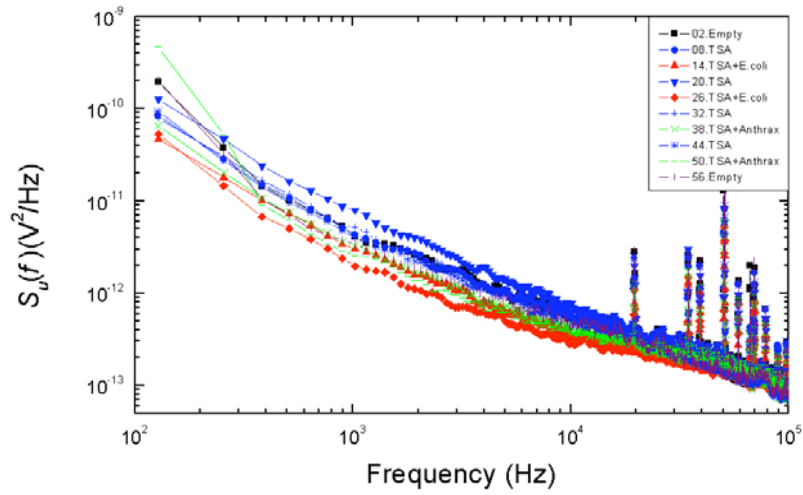

**Figure 7.** Raw Power Spectra of the Heated Sensor TGS 2611. The alias "Anthrax" stands for anthrax surrogate *Bacillus subtilis*.

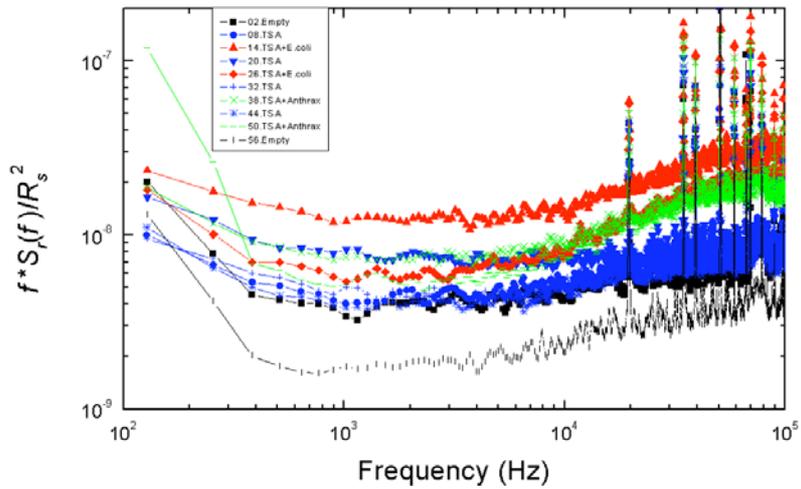

**Figure 8.** Normalized Slope Power Spectra of the Heated Sensor TGS 2611. The alias "Anthrax" stands for anthrax surrogate *Bacillus subtilis*.

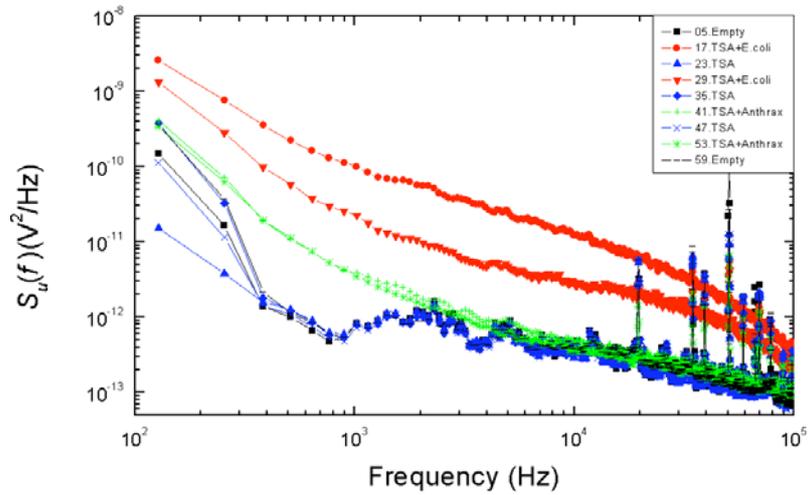

**Figure 9.** Raw Power Spectra of the Sampling-and-hold Sensor TGS 2611. The alias "Anthrax" stands for anthrax surrogate *Bacillus subtilis*.

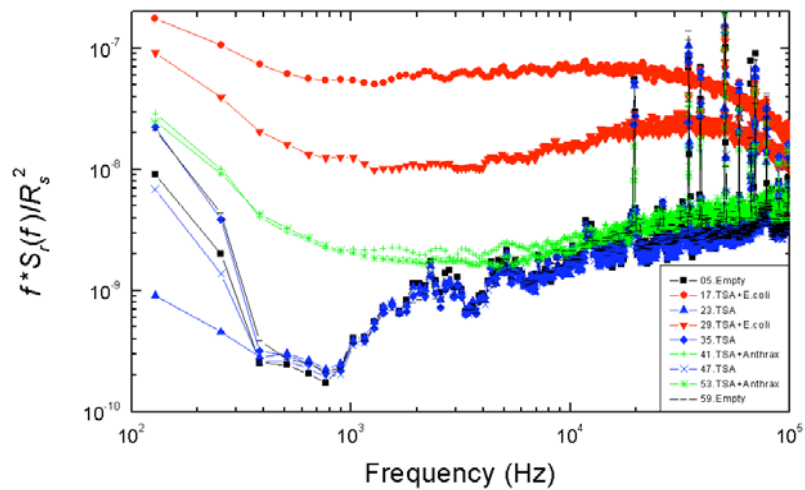

**Figure 10.** Normalized Slope Power Spectra of the Sampling-and-hold Sensor TGS 2611. The alias "Anthrax" stands for anthrax surrogate *Bacillus subtilis*.

In Fig.11, the raw spectra measured with the sensor SP11 in the heated mode are shown. The spectra can be clearly divided into two groups: Empty and TSA; and TSA + Anthrax-surrogate and TSA + E.coli. That means, the empty chamber and the chamber with the TSA are indistinguishable, just like the two bacteria from each other. This is fine for the detection of the presence of bacteria but not for their separate identification. On the other hand, see Fig. 12, the normalized spectra can clearly be divided into four groups, just like the types of the samples, indicating that the normalized spectra of the sensor SP11 in the heated mode can differentiate all samples. This is a further example that the normalization can enhance the selectivity of sensors.

In Fig.13, the spectra measured with the sensor SP11 in the sampling-and-hold mode are shown. The *shapes* of the raw spectral patterns, even though they overlap in several ranges, can clearly be divided into four groups, just like the types of the samples. In Fig.14, the normalized spectra are shown indicating enhanced separation between the four sample types.

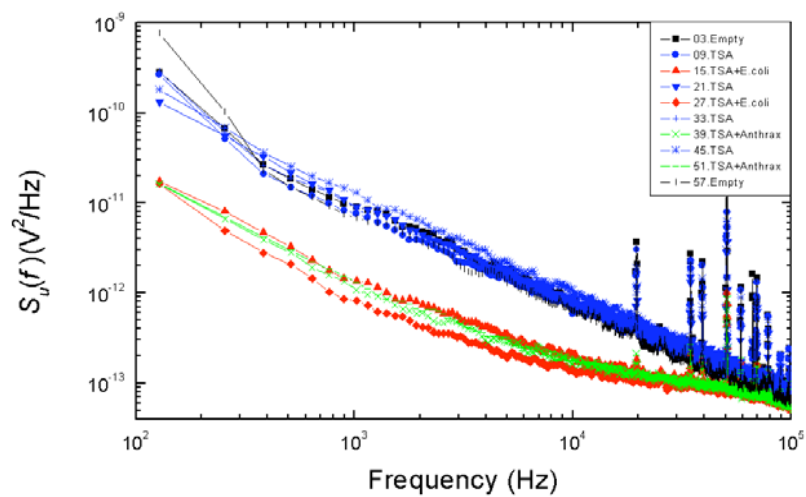

**Figure 11.** Raw Power Spectra of the Heated Sensor SP11. The alias "Anthrax" stands for anthrax surrogate *Bacillus subtilis*.

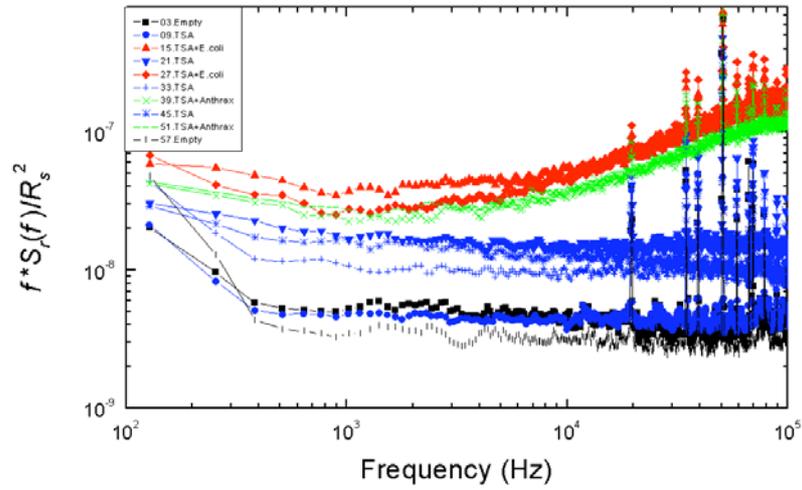

**Figure 12.** Normalized Slope Power Spectra of the Heated Sensor SP11. The alias "Anthrax" stands for anthrax surrogate *Bacillus subtilis*.

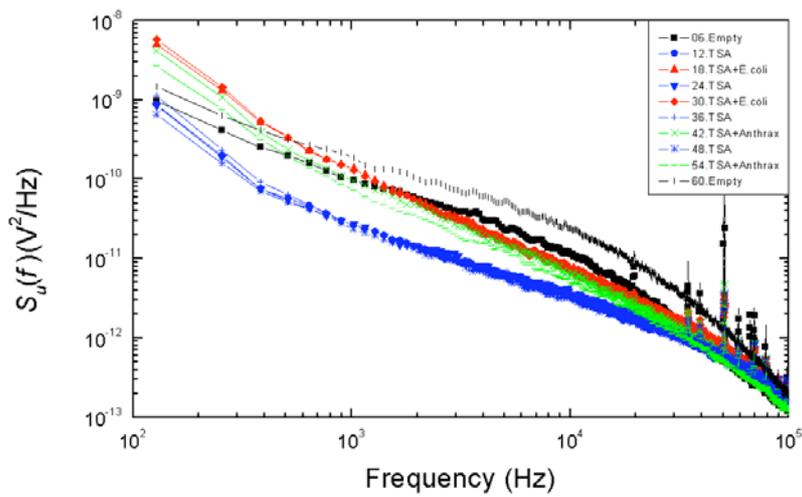

**Figure 13.** Raw Power Spectra of the Sampling-and-hold Sensor SP11. The alias "Anthrax" stands for anthrax surrogate *Bacillus subtilis*.

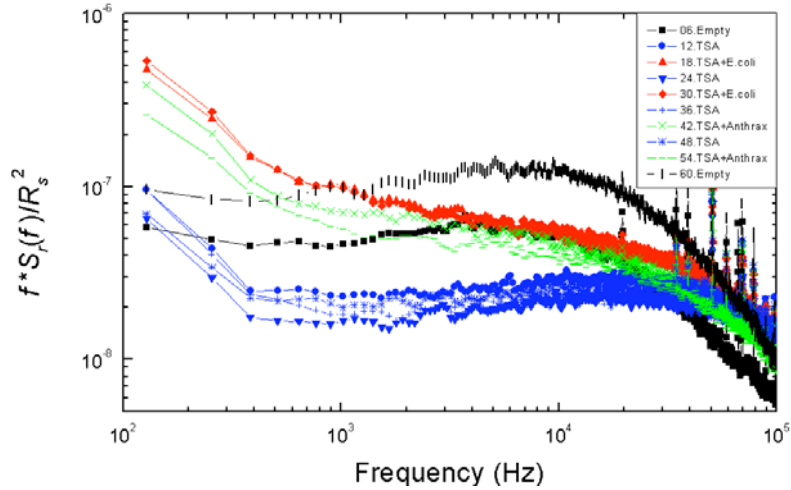

**Figure 14.** Normalized Power Spectra of the Sampling-and-hold Sensor SP11. The alias "Anthrax" stands for anthrax surrogate *Bacillus subtilis*.

## 5. Summary

The conclusions of this exploratory study about the feasibility of using fluctuation-enhanced sensing with single Taguchi type gas sensors to detect and identify different bacteria are summarized in Tables 1 and 2. We note that advanced stochastic signal analysis at the time data level [27] is a powerful tool and has potentials to further enhance the detectability of this type of sensing and will be subject of future studies.

| Sensor | FES Mode | w/o Bacteria | empty/TSA | Bacteria Type |
|---|---|---|---|---|
| SP 32 | Heated | + | + | + |
| SP 32 | Sampling-and-hold | + | + | x |
| TGS 2611 | Heated | ? | x | x |
| TGS 2611 | Sampling-and-hold | + | x | + |
| SP 11 | Heated | + | x | x |
| SP 11 | Sampling-and-hold | + | + | + |

**Table 1**. Summary Table of Raw Spectra: + well detected/identified; x unrecognizable; ? at the limit and it may work fine with advanced pattern recognition.

| Sensor | FES Mode | w/o Bacteria | empty/TSA | Bacteria Type |
|---|---|---|---|---|
| SP 32 | Heated | + | + | + |
| SP 32 | Sampling-and-hold | + | + | + |
| TGS 2611 | Heated | + | + | + |
| TGS 2611 | Sampling-and-hold | + | x | + |
| SP 11 | Heated | + | + | + |
| SP 11 | Sampling-and-hold | + | + | + |

**Table 2:** Summary Table of Normalized Spectra: + well detected/identified; x unrecognizable.

## Acknowledgement

This work was supported in part by the Army Research Office under contract W911NF-08-C-0031.

**Biographies**

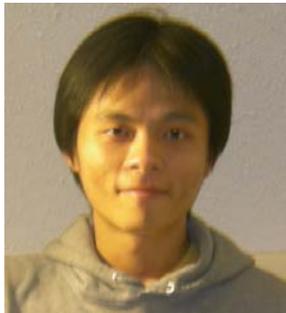**Hung-Chih Chang** was born in Taipei, Taiwan. He received the B.S. degree in electrical engineering from National Central University, Taoyuan, Taiwan in 2000, and the M.S. degree in electrical engineering from National Chiao Tung University, Hsinchu, Taiwan, in 2002. During 2005 and 2006, he worked in TSMC (Taiwan Semiconductor Manufacture Company) in Hsinchu, Taiwan, where he was involved in 90/80/65nm MOS device processes and characterizations. He is currently working towards the Ph.D. degree at the Electrical and Computer Engineering Department of Texas A&M University, College Station. His main research interests include noise and fluctuation, MOSFET modeling, VICOF, chemical and biological sensors.

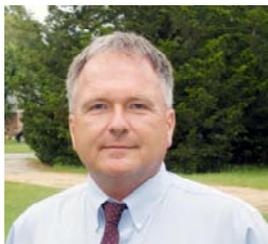**Laszlo B. Kish** (until 1999 Kiss) is full professor at the Electrical and

Computer Engineering Department at Texas A&M University, College Station, USA. PhD (physics, JATE Univ., Hungary, 1984), Docent (solid state physics, Uppsala Univ., Sweden 1994), DSc (physics, Hungarian Acad. of Sci., 2001). His main interests are open questions, especially those related to the laws, limits and applications of stochastic fluctuations, noise. He had been and has been working in many related fields including 1/f noise, stochastic resonance, high-Tc superconductors, chemical and biological sensing, and more recently, the error-speed-energy issues of informatics. He is co-inventor of fluctuation-enhanced sensing, and inventor of the unconditionally secure communication via wire with Johnson-like noise, the zero-power communication, the noise-driven computing scenario and the first noise-based logic scheme. He was the founder Editor-in-Chief of Fluctuation and Noise Letters (2001-2008), and founder of symposium series Fluctuations and Noise (SPIE, 2003) and the conference series Unsolved Problems of Noise (1996).

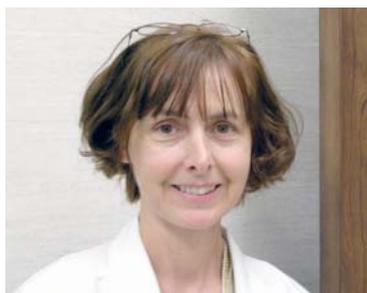**Maria D. King** received her PhD in Microbiology and Chemistry in Germany. She has worked on several microbiology and phage biology related projects that recently led to the development of the prompt bacterial detection technology, the SEPTIC (Sensing of Phage Triggered Ion Cascade). Dr. King is currently conducting biodefense related research studying the effect of bioaerosol collection on the viability and DNA integrity of the aerosolized microorganisms at the Department of Mechanical Engineering, Texas A&M University.

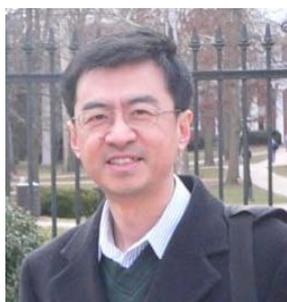**Chiman Kwan** (S'85-M'93-SM'98) received his B.S. degree in electronics with honors from the Chinese University of Hong Kong in 1988 and M.S. and Ph.D. degrees in electrical engineering from the University of Texas at Arlington in 1989 and 1993, respectively.

From April 1991 to February 1994, he worked in the Beam Instrumentation Department of the SSC (Superconducting Super Collider Laboratory) in Dallas, Texas, where he was heavily involved in the modeling, simulation and design of modern digital controllers and signal processing algorithms for the beam control and synchronization system. He received an invention award for his work at SSC. Between March 1994 and June 1995, he joined the Automation and Robotics Research Institute in Fort Worth, where he applied intelligent control methods such as neural networks and fuzzy logic to the control of power systems, robots, and motors. Between July 1995 and March 2006, he was with Intelligent Automation, Inc. in Rockville, Maryland. He served as Principal Investigator/Program Manager for more than sixty five different projects, with total funding exceeding 20 million dollars. Currently, he is the Chief Technology Officer of Signal Processing, Inc., leading research and development efforts in chemical agent detection, biometrics, speech processing, and fault diagnostics and prognostics.

Dr. Kwan's primary research areas include fault detection and isolation, robust and adaptive control methods, signal and image processing, communications, neural networks, and pattern recognition applications. He has published more than 50 papers in archival journals and has had 120 additional papers published in major conference proceedings. He is listed in the New Millennium edition of Who's Who in Science and Engineering and is a member of Tau Beta Pi.